\documentclass{article}
\usepackage{spconf,amsmath,graphicx}

\usepackage{booktabs}
\usepackage{cite}
\usepackage{multirow}
\usepackage{hyperref}

\title{Efficient Training of Self-Supervised Speech Foundation Models\\on a Compute Budget}
\name{Andy T. Liu\textsuperscript{\rm 1,2}, Yi-Cheng Lin\textsuperscript{\rm 1}, Haibin Wu\textsuperscript{\rm 1}, Stefan Winkler\textsuperscript{\rm 2}, Hung-yi Lee\textsuperscript{\rm 1}\thanks{We thank the National Center for High-performance Computing (NCHC) of National Applied Research Laboratories (NARLabs) in Taiwan for providing computational and storage resources.}}
\address{
\textsuperscript{\rm 1}National Taiwan University, Taiwan\\
\textsuperscript{\rm 2}ASUS Intelligent Cloud Services (AICS), Singapore\\
\small\texttt{\{f07942089, r12942075, f07921092, hungyilee\}@ntu.edu.tw}
}
\begin{document}
%
\maketitle
\begin{abstract}
Despite their impressive success, training foundation models remains computationally costly. This paper investigates how to efficiently train speech foundation models with self-supervised learning (SSL) under a limited compute budget. We examine critical factors in SSL that impact the budget, including model architecture, model size, and data size. Our goal is to make analytical steps toward understanding the training dynamics of speech foundation models. We benchmark SSL objectives in an entirely comparable setting and find that other factors contribute more significantly to the success of SSL. Our results show that slimmer model architectures outperform common small architectures under the same compute and parameter budget. We demonstrate that the size of the pre-training data remains crucial, even with data augmentation during SSL training, as performance suffers when iterating over limited data. Finally, we identify a trade-off between model size and data size, highlighting an optimal model size for a given compute budget.
\end{abstract}
\section{Introduction}
\label{sec:intro}

Foundation models, also called upstream models, have gained significant attention in recent years~\cite{mockingjay, npc, wav2vec2, hubert, cpc, modified_cpc, bidir_cpc, apc1, vq_apc, wavlm}. 
There are two stages in the foundation model paradigm~\cite{review, large-scale-eval}:
In the first stage, a pretext learning component, usually a self-supervised learning (SSL) objective, is used to pre-train the foundation model on large amounts of unlabeled data.
In the second stage, these models are adapted to target tasks by training a downstream model~\cite{mockingjay, tera}.
Recently, the best results on speech downstream tasks, such as Automatic Speech Recognition (ASR), often leverage the foundation model paradigm, which involves pre-training models with SSL~\cite{wav2vec2, hubert}.
Foundation models have also showcased achievements of a single universal model capable of delivering promising performance across a wide variety of tasks and domains.~\cite{superb, superb-sg, large-scale-eval}.
Remarkably, this holds true even when presented with a limited amount of task-specific labeled data.

\begin{figure}[tb]
\centering
\includegraphics[width=\linewidth]{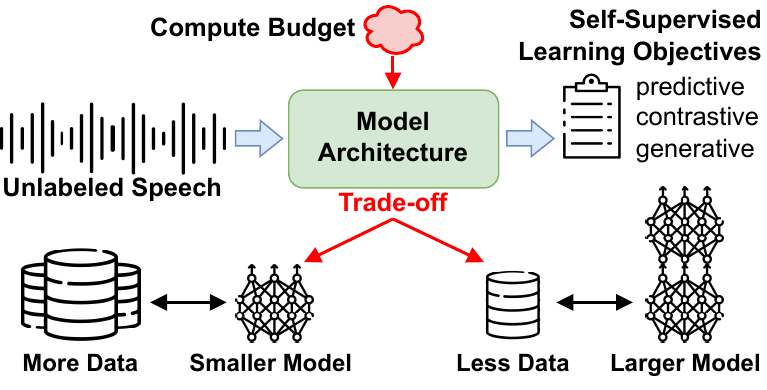}
\caption{We investigate different self-supervised objectives, exploring the trade-offs imposed by computing budgets on model architecture, model size, and data size.
\vspace{-3mm}
}
\label{fig:overview}
\end{figure}

Despite notable advancements in speech with foundation models, the first stage—pre-training models with SSL—requires large memory and high computational costs, making it difficult for many to afford training their own foundation models.
Consequently, the foundation model approach remains unaffordable for many researchers in academia and small companies.
The primary question investigated in this paper is the following: How can we efficiently train speech foundation models under a constrained compute budget?
Instead of adopting or distilling knowledge from an existing foundation model, there are several advantages, especially for groups with limited computing resources, in training their own foundation model.
For example, training a small, targeted, self-supervised foundation model can be beneficial in many domain mismatch scenarios where existing foundation models do not align with the domain of interest.
Even if a foundation model is unavailable for a low-resource domain or language, a substantial amount of unlabeled speech data can still be leveraged.
In this scenario, one could first pre-train a foundation model using the unlabeled data before fine-tuning it for the desired downstream task.

However, recent research efforts in speech have primarily focused on crafting new algorithms, enhancing self-supervised objectives, or adapting existing methods to unexplored situations.
Only a few studies have focused on the goal of training SSL models efficiently within a compute budget~\cite{distilhubert, melhubert, on-device-ssl, sustainable-ssl}.
Furthermore, the specific factors contributing to the success of speech foundation models remain inconclusive, and the training dynamics for efficiency are still not well understood.
Although there is extensive research on resource-efficient training in the context of Large Language Models (LLMs)~\cite{optimal-llm}, the findings may not be directly applicable to speech, as speech inputs are often substantially longer than text.
This type of work is still lacking in the speech domain, particularly in the context of speech foundation models.

Our work aims to bridge the current research gap by investigating the key factors for resource-efficient pre-training speech foundation models within a compute budget.
Figure~\ref{fig:overview} provides a comprehensive overview of the factors investigated in this work.
It illustrates a practical situation where, with a limited compute budget, a trade-off arises between model architecture, model size, and data size.
We pose the following research questions, which we will answer through our experiments, bearing in mind the limitations imposed by a restriction in compute budget:
1) How does the choice of self-supervised objectives affect foundation models' performance?
2) To what extent does the model architecture contribute to foundation models' performance?
3) How does the size of pre-training data influence the performance of foundation models, particularly in the context of iterating over a small data size compared to a larger one?
4) Given a fixed computational budget, is there an ideal model size for foundation models that can maximize performance?
To the best of our knowledge, these research questions have not yet been answered or verified for speech SSL.

To address these research questions, we consider self-supervised learning objectives categorized as predictive, contrastive, and generative, as described in a recent survey paper \cite{review}.
We then systematically examine foundation models trained with these objectives under a compute budget.
By isolating key pre-training factors one at a time, we investigate how model architecture, model size, and data size impact the final performance.
We adopt the SUPERB~\cite{superb} benchmark for downstream evaluation of our foundation models, ensuring comprehensive and reproducible results.
Although we use SUPERB as a benchmark, the goal of this paper is not to outperform the existing scores on the benchmark but to make analytical steps toward fully understanding the training dynamics of self-supervised foundation models.
Additionally, we carefully discuss the challenges and considerations posed by limited computational resources.
Our work provides the following insights and contributions, addressing the previously listed research questions:
\begin{enumerate}
\item We compare self-supervised objectives within a fully standardized and controlled setup, offering a unique contribution to the field. Our findings indicate that SSL objectives can influence the performance of foundation models. However, their impact is not as significant as other key pre-training factors. (Section~\ref{ssec:ssl-objectives})
\item With the same compute and parameter budget, slimmer SSL models have been observed to outperform the three-layer small SSL models commonly used in previous work~\cite{tera, npc, apc1, vq_apc, audioalbert, distilhubert, on-device-ssl}. (Section~\ref{ssec:architecture})
\item We investigate the trade-off between data size and iterations per data point. Our findings show that increasing the volume of data for pre-training is beneficial, but more importantly, the pre-training data size must be sufficiently large. (Section~\ref{ssec:data-size})
\item We demonstrate U-shaped performance curves, indicating that there is an optimal model size for pre-training within a given compute budget. This suggests that a balance between computational resources and model performance can be achieved. (Section~\ref{ssec:valley-curve})
\end{enumerate}

\section{Related Work}
\label{sec:related}
\subsection{Lowering Computational Costs for Speech SSL}
Several existing works explore the issue of high computational costs for speech SSL~\cite{sustainable-ssl}.
Some methods use knowledge distillation~\cite{distilhubert, fithubert} or pruning~\cite{prune-for-ssl-asr, structured-prune-for-ssl}, but these rely on a fully pre-trained large model, as they cannot be trained independently.
The work of MelHuBERT~\cite{melhubert} replaces the convolutional module in HuBERT with Mel Spectrograms to reduce compute requirements and pre-training time.
There are also efforts to make wav2vec 2.0 more compute-efficient, including squeezing the input~\cite{compute-reduction-w2v2, performance-efficient-w2v2} or truncating the input~\cite{match-to-win}.
In contrast to prior work that primarily focused on reducing computing costs for a single SSL method, our work explores a different angle by identifying the impact of key components on computational costs at a more fundamental level while being able to complement existing techniques.

\subsection{Network Architectures for Speech SSL}
Some existing work explores the effects of different architectures for SSL, but most studies focus on introducing new architectures without considering the trade-offs between model width and depth.
Our experiments demonstrate that these trade-offs are important for SSL.
A previous study~\cite{similarity} examined the impact of model building blocks such as RNN, Transformer, and CNN.
However, their investigation was primarily based on the measurement of representation similarity.
In addition, they did not conduct experiments under a fixed compute budget.
In \cite{deep-pre-training}, the authors demonstrate that for computer vision, combining depth with pre-training provides a good prior for model weights.
However, this has not been verified for recent pre-training methods on speech, such as self-supervised foundation models.

\subsection{Scaling Model and Data Sizes in Speech SSL}
Some existing work seeks to improve performance through scaling, but most focus on adding more data to larger models without considering the associated costs and trade-offs in training.
These factors are crucial for the affordability of SSL.
In \cite{bidir_cpc, wavlm}, the authors reported enhanced downstream performance due to the use of larger datasets.
However, these studies focused solely on performance improvements from adding more data and did not investigate the trade-offs between model size and data size.
The study presented in \cite{scaling-effect} explored the scaling effect between model sizes and performances.
The authors attempted to establish a relationship between model size and self-supervised L1 loss, demonstrating that the relationship follows a power law approximately.
However, they trained their models with a constant total step across all model sizes and did not account for the additional resources required for larger models.


\subsection{Analytical Approaches for Speech SSL}
Recent research efforts have also focused on various analytical aspects, including self-attention mechanisms~\cite{understanding}, resilience to biases~\cite{data-bias, social-bias}, and the encoding of different types of information~\cite{what_ssl_encode}.
The paper \cite{ssl-more-phonetic} reveals that SSL representations consistently and significantly exhibit more phonetic than semantic similarity.
In \cite{ssl_orthogonal}, the authors investigate how self-supervised speech representations distribute speaker and phonetic information, concluding that these are encoded in nearly orthogonal subspaces.
Unlike previous work, our analytical study approaches the essence of SSL from a different angle.
We simultaneously consider several key factors in pre-training SSL models, including the self-supervised objective, model architecture, model size, and data size, all within a compute budget constraint.
We also provide a more comprehensive view by experimenting with a large set of downstream evaluation tasks.

\section{Experimental Setup}
\label{sec:exp-setup}

\subsection{Selection of Self-Supervised Objectives}
The prevailing trend in the field of speech foundation models focuses on pre-training with self-supervised objectives.
In recent work~\cite{review}, SSL models are categorized into predictive, contrastive, or generative based on the nature of their respective self-supervised objectives.
Therefore, we carefully select representative objectives from these three categories, specifically HuBERT~\cite{hubert} (predictive), wav2vec~2.0~\cite{wav2vec2} (contrastive), and TERA~\cite{tera} (generative).
Conveniently, these three objectives can be standardized with identical model components and trained using the same toolkit.
This standardization allows us to use the same model architecture across different self-supervised learning (SSL) objectives, a comparison not previously examined in the literature.
We minimize potential confounding factors that could influence final performance by pre-training various SSL objectives with consistent building blocks.
In this paper, we construct all models using consistent components, including a convolutional encoder, Transformer encoder blocks, and a projection layer.
We implement and train these models using the Fairseq\footnote{\href{https://github.com/facebookresearch/fairseq}{https://github.com/facebookresearch/fairseq} \label{fairseq}}\footnote{\href{https://github.com/andi611/fairseq/tree/master/examples/tera}{https://github.com/andi611/fairseq/tree/master/examples/tera} \label{fairseq-tera}} toolkit.

\subsection{Model Architecture and Model Size}
When computational resources are limited, researchers often resort to smaller configurations for SSL models.
This approach is exemplified by the model architectures used in previous works such as TERA~\cite{tera}, NPC~\cite{npc}, APC~\cite{apc1}, VQ-APC~\cite{vq_apc}, Audio ALBERT~\cite{audioalbert}, DistilHuBERT~\cite{distilhubert}, and the student model in \cite{on-device-ssl}, which all used models with three layers or fewer.
However, as our results will show, this common choice might not be optimal.
In our experiments, we first establish a 3-layer \textit{Small} model, following the above literature, which results in approximately 20 million parameters.

On the other hand, we propose a different small model, denoted as \textit{Slim}.
While \textit{Slim} models have the same parameter size as \textit{Small} models, they feature a narrower width and a greater depth of 12 layers.
We use these two settings to explore the effect of different model architectures under the same compute and parameter budget.
We also investigate smaller model sizes (30\%, 50\%, and 70\% of \textit{Slim}) and larger model sizes (200\%, as well as the \textit{Base} 476\% and \textit{Large} 1590\% models described in \cite{hubert, wav2vec2}).
We list the model details in Table~\ref{table:model-details}.
These specific model sizes are used to explore the trade-off between model size and data size in our subsequent experiments.
The goal of these settings is not to suggest the optimal hyperparameters for best performance, but to make reasonable changes in the hyperparameters to demonstrate a U-shaped performance curve (Fig~\ref{fig:valley-curve}).

\subsection{Self-Supervised Pre-training Setup}
Following \cite{hubert, wav2vec2, tera}, we use the LibriSpeech dataset~\cite{librispeech}, which provides up to 960 hours of speech, to pre-train all our foundation models.
All models take speech waveforms sampled at 16kHz as input.
We use the Adam optimizer~\cite{Adam} with a learning rate of \textit{5e-4}, as outlined in \cite{hubert, wav2vec2}, and a batch size of 87.5 seconds of audio per GPU, following \cite{hubert}.
For HuBERT models, we use the default cluster size and follow the iterative clustering process described in \cite{hubert}.
We compute training FLOPS (floating-point operations per second) as described in \cite{optimal-llm}, implemented with the DeepSpeed\footnote{\href{https://github.com/microsoft/DeepSpeed}{https://github.com/microsoft/DeepSpeed} \label{DeepSpeed}} FLOPS profiler.
We set the final training FLOPS at $1.33 \times 10^{18}$ for our experiments, which is the amount required to train a \textit{Slim} model for 400k steps, taking approximately three days on two GPUs.
The FLOPs budget and 400k steps were chosen based on preliminary experiments to balance training time and model convergence.
To ensure fair comparisons, other pre-training configurations are identical to the original work~\cite{hubert, wav2vec2, tera}.

\subsection{Downstream Evaluation Methods}
We evaluate our pre-trained foundation models using tasks from the widely recognized SUPERB Benchmark~\cite{superb}.
The downstream tasks are listed in Table~\ref{table:superb}.
These tasks are sourced from six different downstream datasets, as described in \cite{superb}.
Following the SUPERB challenge protocol, we freeze our foundation models and do not fine-tune them with the downstream model.
Following the SUPERB paper, to accurately capture each SSL model’s performance, we sweep the optimal learning rate from \textit{1e-1} to \textit{1e-5} on a log scale for each combination of foundation model and downstream task~\cite{superb}.
We follow all standards of SUPERB for our downstream evaluation to allow easy comparison with other results.
The benchmarking scripts are sourced from the S3PRL toolkit\footnote{\href{https://github.com/s3prl/s3prl}{https://github.com/s3prl/s3prl} \label{s3prl}}. 

\begin{table}[h]
\centering
\setlength{\tabcolsep}{3pt}
\resizebox{\columnwidth}{!}{
\begin{tabular}{l|l}
\toprule
Downstream Tasks & Evaluation Metrics \\
\midrule
\textbf{ASR} (Auto. Speech Recognition) & word error rate (WER) $\downarrow$ \\
\textbf{PR} (Phoneme Recognition) & phone error rate (PER) $\downarrow$ \\
\textbf{ASV} (Auto. Speaker Verification) & equal error rate (EER) $\downarrow$ \\
\textbf{SD} (Speaker Diarization) & diarization error rate (DER) $\downarrow$ \\
\textbf{SF} (Slot Filling) & slot value CER $\downarrow$ \\
\textbf{SF} (Slot Filling) & slot-type F1 score $\uparrow$ \\
\textbf{KS} (Keyword Spotting) & accuracy (ACC) $\uparrow$ \\
\textbf{IC} (Intent Classification) & accuracy (ACC) $\uparrow$ \\
\textbf{SID} (Speaker Identification) & accuracy (ACC) $\uparrow$ \\
\textbf{ER} (Emotion Recognition) & accuracy (ACC) $\uparrow$ \\
\bottomrule
\end{tabular}}
\caption{Downstream tasks and metrics from SUPERB. The symbol $\downarrow$ indicates a lower score is better, vice-versa for $\uparrow$.
\vspace{-3mm}
}
\label{table:superb}
\end{table}
\begin{table*}[ht]
\small
\centering
\begin{tabular}{|ll||c|c|c|c|cc|c|c|c|c|}
\hline
\multirow{2}{*}{SSL Objective} & \multirow{2}{*}{Arch.} & ASR & PR & ASV & SD & \multicolumn{2}{c|}{SF} & KS & IC & SID & ER \\ \cline{3-12}

& & WER $\downarrow$ & PER $\downarrow$ & EER $\downarrow$ & DER $\downarrow$ & CER $\downarrow$ & F1 $\uparrow$ & ACC $\uparrow$ & ACC $\uparrow$ & ACC $\uparrow$ & ACC $\uparrow$ \\ \hline \hline

\multirow{2}{*}{HuBERT~\cite{hubert}} & \textit{Small} & 19.37 & 36.30 & 8.82 & 8.04 & 50.46 & 70.28 & 89.39 & 63.09 & 59.12 & 59.38 \\ \cline{2-12}
& \textit{\textbf{Slim}} & \textbf{14.56} & \textbf{21.85} & \textbf{7.15} & \textbf{7.16} & \textbf{35.44} & \textbf{82.04} & \textbf{93.51} & \textbf{84.31} & \textbf{59.19} & \textbf{60.34} \\ \hline \hline
\multirow{2}{*}{wav2vec 2.0~\cite{wav2vec2}} & \textit{Small} & 20.18 & 37.38 & 10.49 & 9.67 & 48.93 & 72.48 & 89.45 & 65.36 & 43.81 & 59.93 \\ \cline{2-12}
& \textit{\textbf{Slim}} & 17.01 & 27.66 & 8.93 & 8.20 & 40.83 & 78.34 & 91.27 & 74.14 & 40.46 & 60.14 \\ \hline \hline
\multirow{2}{*}{TERA~\cite{tera}} & \textit{Small} & 19.43 & 36.97 & 9.97 & 8.53 & 52.04 & 70.11 & 87.37 & 52.07 & 53.21 & 58.41 \\ \cline{2-12}
& \textit{\textbf{Slim}} & 16.26 & 30.47 & 7.61 & 10.09 & 43.88 & 75.08 & 91.50 & 66.36 & 55.14 & 59.05 \\ \hline
\end{tabular}
\caption{Comparison of self-supervised objectives, \textit{Small} and \textit{Slim} models, under a fixed compute and parameter budget.}
\label{table:small-slim}
\end{table*}
\begin{table*}[ht]
\small
\centering
\begin{tabular}{|c||c|c|c|c|cc|c|c|c|c|}
\hline
\multirow{2}{*}{Data Size} & ASR & PR & ASV & SD & \multicolumn{2}{c|}{SF} & KS & IC & SID & ER \\ \cline{2-11}

& WER $\downarrow$ & PER $\downarrow$ & EER $\downarrow$ & DER $\downarrow$ & CER $\downarrow$ & F1 $\uparrow$ & ACC $\uparrow$ & ACC $\uparrow$ & ACC $\uparrow$ & ACC $\uparrow$ \\ \hline \hline

1 hours & 39.60 & 68.05 & 21.56 & 10.49 & 51.59 & 72.02 & 70.20 & 30.32 & 10.21 & 53.23 \\ \hline
10 hours & 23.88 & 46.62 & 12.52 & 9.25 & 47.48 & 74.41 & 84.13 & 48.62 & 30.37 & 55.03 \\ \hline
100 hours & 18.66 & 28.59 & 9.69 & 8.67 & 42.43 & \textbf{78.83} & 89.87 & 63.09 & 38.90 & \textbf{60.29} \\ \hline
\textbf{960 hours} & \textbf{17.01} & \textbf{27.66} & \textbf{8.93} & \textbf{8.20} & \textbf{40.83} & 78.34 & \textbf{91.27} & \textbf{74.14} & \textbf{40.46} & 60.14 \\ \hline
\end{tabular}
\caption{Comparison of different data sizes and data iterations, with training steps adjusted to ensure constant final training FLOPS. Due to space limitations, the presented results are for wav2vec 2.0 \textit{Slim}. HuBERT and TERA exhibit similar trends.}
\label{table:data-size}
\end{table*}
\begin{table*}
\centering
\small 
\setlength{\tabcolsep}{3pt}
\begin{tabular}{cc|ccc|cccc}
\toprule
& relative & & number of parameters & &  \multicolumn{4}{c}{transformer hyperparameter} \\
Size & to \textit{Slim} & HuBERT  & wav2vec 2.0 & TERA  & n\_layers & d\_model & ffw\_size & n\_heads \\
\midrule
& 30\% & 5.9M & 6.2M & 5.7M & 4 & 192 & 320 & 4 \\
& 50\% & 9.9M & 10.3M & 9.8M & 5 & 320 & 768 & 8 \\
& 70\% & 14.1M & 14.4M & 13.9M & 7 & 384 & 768 & 8 \\
\textit{Small} & 100\% &  20.9M & 21.3M & 20.7M & 3 & 640 & 2048 & 8 \\
\textit{Slim}  & 100\% & 20.0M & 20.4M & 19.8M & 12 & 384 & 768 & 8 \\
& 200\% & 40.0M & 40.4M & 39.8M & 15 & 528 & 1024 & 12 \\
\textit{Base}  & 467-476\% & 94.7M & 95.0M & 94.5M & 12 & 768 & 3072 & 12 \\
\textit{Large} & 1559-1590\% & 316.6M & 317.4M & 315.6M & 24 & 1024 & 4096 & 16 \\
\bottomrule
\end{tabular}
\caption{Details of different model architectures and sizes for HuBERT, wav2vec 2.0, and TERA.}
\label{table:model-details}
\end{table*}
\begin{figure*}[htb]
\centering
\includegraphics[width=\linewidth]{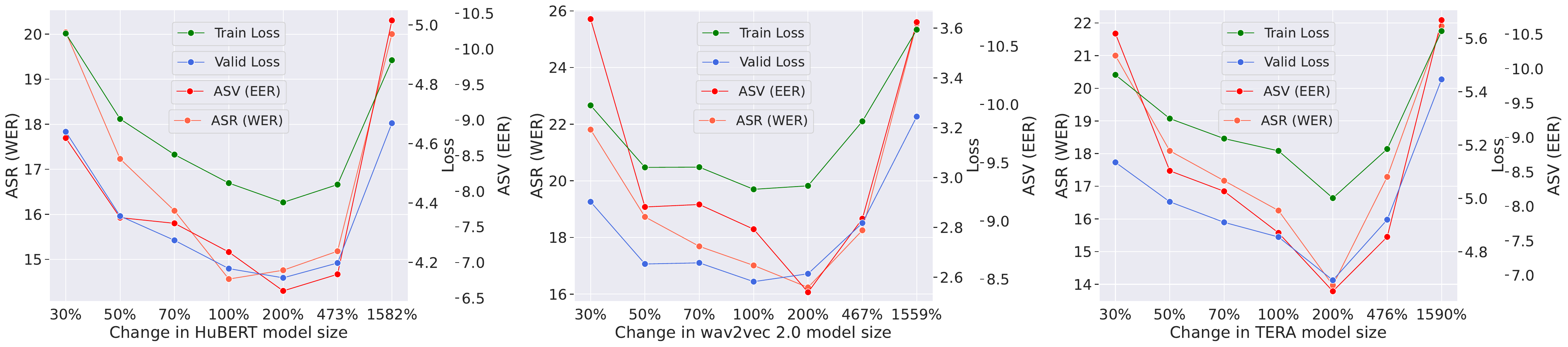}
\caption{The U-shaped performance curves illustrate the trade-off between model size and data size, with training FLOPS kept constant across all data points. The U-shaped curves suggest the existence of an optimal model size for a given compute budget.}
\label{fig:valley-curve}
\end{figure*}

\section{Results and Analysis}
\label{sec:result}

\subsection{The Effect of Self-Supervised Learning Objectives}
\label{ssec:ssl-objectives}
While we use SUPERB to benchmark our SSL models, our aim is not to compare them against state-of-the-art (SOTA) models on the SUPERB leaderboard or to determine the best SSL objective.
Instead, our goal is to compare SSL objectives within a fully controlled setup to assess their contributions to the success of SSL.
The existing literature~\cite{superb, superb-sg, review} does not offer this insight, as previous work only constrains the downstream model, not the SSL models.
For instance, the building recipes for each SSL model on the SUPERB leaderboard vary, making it difficult to isolate and compare individual factors like the self-supervised learning objective.

Table~\ref{table:small-slim} shows a comparative evaluation of the predictive, contrastive, and generative self-supervised objectives, as represented by the HuBERT, wav2vec~2.0, and TERA models, respectively.
For this experiment, we set the pre-training data size to 960 hours, and all models are operated within identical computational and parameter constraints.
We observed that HuBERT consistently surpasses wav2vec~2.0 and TERA in performance across all downstream tasks.
Conversely, wav2vec~2.0 and TERA display variability in their performance, each exhibiting strengths and weaknesses for different tasks when compared to one another.
The above observations persist across both the \textit{Slim} and \textit{Small} model architectures, a topic we will explore further in Section~\ref{ssec:architecture}.
We conclude that within computational and parameter budgets, the choice of self-supervised objective does influence downstream performance to some extent.

\subsection{\textit{Small} vs. \textit{Slim} Model Architectures}
\label{ssec:architecture}
In this section, we isolate and compare the factor of model architecture by fixing the pre-training data size to 960 hours and ensuring all models adhere to a fixed compute and parameter budget.
Table~\ref{table:small-slim} compares \textit{Small} and \textit{Slim} models for all self-supervised objectives.
The \textit{Slim} models outperform their \textit{Small} counterparts for all objectives.
Our findings suggest a potentially more effective alternative in terms of model architecture design.
We validate this observation across HuBERT, wav2vec~2.0, and TERA, observing consistent improvements on the \textit{Slim} version over \textit{Small}.
Our results imply that many previously proposed small models~\cite{tera, npc, apc1, vq_apc, audioalbert, distilhubert, on-device-ssl} could benefit from a narrower and deeper build.
Finally, although the HuBERT objective generally outperforms other methods, we note that the choice of architecture has a more significant impact on performance than the self-supervised objective.

\begin{table}[ht]
  \centering
  \resizebox{\columnwidth}{!}{
  \begin{tabular}{llllll}
    \toprule
    SSL & Arch. & Relative & \# Params & ASR & ASV \\
    Objective & & to \textit{Small} & & WER$\downarrow$ & EER$\downarrow$ \\
    \midrule \midrule
    \multirow{2}{*}{HuBERT} & \textit{Small} & 100\% & 20.9M & 19.37 & 8.82 \\
    & \textit{\textbf{Slim}} & \textbf{200\%} & 40.0M & \textbf{14.76} & \textbf{6.60} \\
    \midrule
    \multirow{2}{*}{wav2vec~2.0} & \textit{Small} & 100\% & 21.3M & 20.18 & 10.49 \\
    & \textit{\textbf{Slim}} & \textbf{200\%} & 40.4M & \textbf{16.24} & \textbf{8.39} \\
    \midrule
    \multirow{2}{*}{TERA} & \textit{Small} & 100\% & 20.7M & 19.43 & 9.97 \\
    & \textit{\textbf{Slim}} & \textbf{200\%} & 39.8M & \textbf{13.97} & \textbf{6.77} \\
    \bottomrule
  \end{tabular}}
\caption{Summary of our findings on improving conventional \textit{Small} models within a budget. All models are trained under the same computing budget.}
\label{table:improved-models}
\end{table}

\subsection{Trade-off Between Data Size and Data Iteration}
\label{ssec:data-size}
In this section, we first isolate and compare the impact of data size while fixing the self-supervised learning objective, model size, and compute budget.
Table~\ref{table:data-size} presents our measurements for boosting the pre-training unlabeled data size from 1 hour to 960 hours.
Due to space limitations, we present the results of wav2vec~2.0 \textit{Slim}.
With the same experiment settings, HuBERT and TERA show identical trends.
We compare the difference of pre-training on 100 hours and 960 hours of speech (Table~\ref{table:data-size}, rows three and four).
Despite a nearly tenfold increase in pre-training data size, we observe limited performance improvements.

Next, we investigate the issue of data iteration.
Within a fixed computational budget, there is a trade-off between data size and data iteration.
One option is to train the model on a larger dataset (more diversity) with fewer updates per data point (less data iteration).
Alternatively, we could prioritize more frequent updates on a smaller dataset, allowing the model to learn repeatedly from the same data but reducing the overall diversity of the data it is exposed to.
When trading off between data size and data iteration, empirically we do not observe overfitting due to the strong data augmentation in all self-supervised objectives.
Therefore, we do not use an early stopping criterion during pre-training.
Instead, we select the final pre-training iteration to fully utilize the available training FLOPS.
As we decrease the data size, we allow the models to receive more frequent updates on the same data with dynamic data augmentation from the self-supervised objective.
In Table~\ref{table:data-size}, we experiment with varying sizes of unlabeled data, reducing from an initial 960 hours to 100, 10, and finally, 1 hour.
The FLOP budget remains constant for all settings.

Our experimental results show a decline in model performance when the pre-training data size is reduced from 960 to 100 hours (rows three and four).
The performance degradation becomes significantly pronounced when the pre-training data size drops below 100 hours (rows one and two).
This finding suggests that pre-training data size has more influence on the performance of foundation models than the number of iterations, particularly when the data size drops below a certain point. 
Investing in new, diverse data holds greater significance than simply revisiting the same data multiple times.
Interestingly, SSL foundation models often undergo data augmentation, like masking, as part of their pre-training task.
This suggests that the quality of unlabeled data plays a significant role in pre-training SSL foundation models.

\subsection{Trade-off Between Model Size and Data Size}
\label{ssec:valley-curve}

When operating within a compute budget, there is a necessary trade-off between the size of the model and the pre-training data size.
In Figure~\ref{fig:valley-curve}, we vary the model sizes while maintaining consistent training FLOPS.
The model details are listed in Table~\ref{table:model-details}.
Note that these hyperparameters are used solely to demonstrate the U-shaped performance curves for each objective and do not represent the optimal settings for best performance.
All models are pre-trained with a data size of 960 hours to ensure maximum data diversity.
Due to constrained FLOPS, the number of update steps decreases as we scale up the model's size. 
This implies that larger models have access to less data iteration throughout their training process.
The final FLOPS allowance enables the largest model to iterate for approximately 3.25 epochs, while the smallest model completes around 24.74 epochs.
This experimental design enables us to answer the question: Is there an optimal model size for a given FLOPS budget?

We assess each model's smoothed training loss and validation loss (on LibriSpeech dev-clean) and the downstream performance on ASV and ASR.
For all metrics, a lower score means better performance ($\downarrow$).
The results are shown in Figure~\ref{fig:valley-curve}, where 100\% corresponds to our \textit{Slim} model.
Our findings suggest that for HuBERT, wav2vec~2.0, and TERA, the most efficient model size for training within the predetermined FLOPS budget tends to align with approximately twice the original model size (200\% of \textit{Slim}).
Our results illustrate the necessary trade-offs between model and data size when dealing with a limited compute budget.
We show that an optimal model size exists for efficient training for a given FLOPS budget.
This implies that most of the current SSL models may not be optimal in size and can still be improved.

It is crucial to highlight that merely increasing the model size, without a corresponding increase in the compute budget, does not automatically translate to better performance.
This is evidenced in Figure~\ref{fig:valley-curve}, where larger models (\textit{Base} and \textit{Large)}, constrained by the same compute budget, tend to underperform.
This observation reinforces the importance of a balanced approach to model scaling within the limits of available computational resources, ensuring that increases in model size are meaningfully aligned with the objective of optimizing performance.
Furthermore, we point out that all self-supervised objectives exhibit a strong correlation between pre-training loss, ASV, and ASR performance, as all the metrics show similar trends.

\subsection{Improving Small Models Under a Compute Budget}
\label{ssec:improved-models}

Table~\ref{table:improved-models} summarizes the results of resource-efficient pre-training on a budget, based on our previous findings.
When computing resources are limited, researchers often use a setup similar to the \textit{Small} model~\cite{tera, npc, apc1, vq_apc, audioalbert, distilhubert, on-device-ssl}.
By combining the \textit{Slim} architecture with an optimal model size of 200\%, we improve over the performance of conventional \textit{Small} models.
We also point out that all self-supervised objectives can be improved using our general findings, all within the same compute budget.
Note that the compute budget remains consistent for both the 100\% and 200\% model sizes by reducing the training time for the larger models to maintain computational parity.
Interestingly, after improvement, the TERA objective outperformed the other two objectives, overturning the initial \textit{Small} setup ranking.
However, as noted in previous findings, the aim of this paper is not to determine the best SSL objective or to achieve new SOTA results, but to provide insights for efficient SSL training on a budget.

\section{Conclusion}
\label{sec:conclusion}
We believe pre-training speech foundation models should be affordable for the many.
In this work, we offer insights into the training dynamics of SSL speech foundation models under compute budget constraints.
We find that factors beyond the SSL objectives significantly influence the success of SSL.
We identify that model architecture profoundly impacts performance and show a trade-off between data size and data iterations.
While more pre-training data is generally beneficial, having sufficient data is essential.
Additionally, we identify an optimal model size for a given compute budget, indicating a balance between budget and performance.
Our research offers guidance for future resource-efficient model pre-training in compute-constrained scenarios. 



\bibliographystyle{IEEEbib}
\bibliography{refs}

\end{document}